\documentclass[a4paper]{article}

\usepackage[T1]{fontenc}                       
\usepackage[utf8]{inputenc}                    %
\usepackage[english]{babel}                    %

\usepackage[bitstream-charter]{mathdesign}     
\usepackage[vvarbb,bigdelims]{newtxmath}

\usepackage{relatese}
\usepackage{tabularx}
\usepackage{booktabs}

\hypersetup{
  pdfinfo = {
    Author = {Francisco Mariano-Neto, Thiago de Castro Pereira},
    Title = {Recent Nanoninformatics Approaches for Developments in Nanobiotechnology and Nanomedicine},
    Subject = {Nanoinformatics techniques in Medical Physics},
    Keywords = {nanoinformatics, nanomedicine, nanobiotechnology},
  }
}

\title{Recent Nanoninformatics Approaches for Developments in Nanobiotechnology and Nanomedicine}
\author[1]{Francisco Mariano-Neto}
\author[1]{Thiago de Castro Pereira}

\affil[1]{Instituto de Física - Universidade de São Paulo}
\affil[2]{Faculdade de Medicina - Universidade de São Paulo}

\begin{document}

\maketitle

\section{Introduction}

\subsection{Nanobiotechnology and Nanomedicine}

Nanobiotechnology is a highly convoluted interdisciplinary field that conjoins several different areas of research, like Physics, Chemistry, Biology and Engineering. It is futile to try and pinpoint a start date for this field, because it has been intertwined into human activity from very early. Evidence of the usage of nanomaterials suggests the practice predates the 4th century AD\cite{kaurChapterNanobiotechnologyMethods2021}. Before that, uses of gold nanoparticles in ancient Indian medicines have been documented\cite{brownNanogoldpharmaceutics2007}.

While no precise date can be determined for when the field effectively came into existence, it can be argued that one of the first advocates for the potential of deliberate application of the properties arising from the nanometric scale structure of materials was Richard P. Feynman, in a widely-known talk from 1960 titled "There is plenty of room at the bottom"\cite{feynman1960there,feynmanPleasureFindingThings2000,heyFeynmanComputationExploring2018}.

Following that, the inexorable evolution of systematic research in this domain has led to a wide spectrum of applications. By 2015, there were already 1814 consumer products cataloged\cite{vanceNanotechnologyRealWorld2015}. Research of nanotechnology in life sciences has been particularly prolific. Owing to the similarity in scale between nanomaterials and biologic molecules and systems, the investigation of nanotechnology to medicine led to a prompt emergence of publications. This interface between medical research and nanotechnology, referred to as Nanomedicine\cite{kimNanomedicine2010}, grew very rapidly and currently comprises about half of all the publications related to nanotechnology. Figure \ref{fig:ratio} below shows the number of articles per year, according to a search for the MeSH keywords “Nanotechnology" and “Nanomedicine", with the yellow curve showing the proportion of works related to nanotechnology that also fall under the "nanomedicine" MeSH keyword.

\begin{figure}[H]
	\centering
	\includegraphics[width=.7\linewidth]{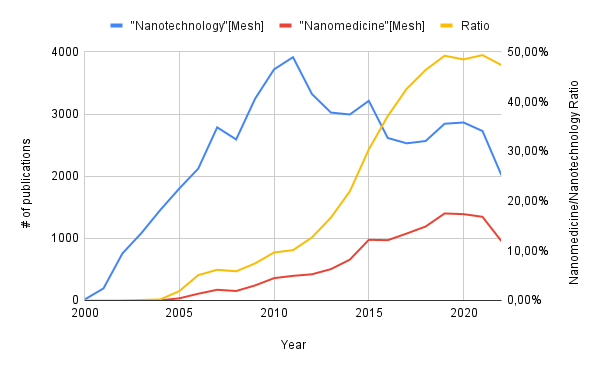}
	\caption{Evolution of articles related to "nanomedicine" and "nanotechnology", as well as the ratio between them \label{fig:ratio}}
\end{figure}

Physical methods for synthesis of nanoparticles follow a "top-down" approach, in which the bulk material is processed to then generate smaller and smaller fragments, reaching the nanometric scale. Chemical processes, on the other hand, follow a "bottom-up" approach, during which compounds are reduced and cluster to form particles. Biological pathways employ cells and biological substances for reduction to produce mainly metallic nanoparticles\cite{kaurChapterNanobiotechnologyMethods2021}.

The participation of informatics-related disciplines is not new to research in life sciences. The contributions of computer scientists have been a staple in the field, from day-to-day operations\cite{maojoMedicalInformaticsBioinformatics2001} to groundbreaking research such as the Human Genome Project\cite{NewGoalsHuman}. Nevertheless, being at the bleeding edge of scientific research, nanobiotechnology has imposed several challenges in terms of the determination of the physicochemical properties of nanoparticles, pushing against the limits of several widely used techniques\cite{gonzalez-niloNanoinformaticsEmergingArea2011}.

Emerging from the synergy between bioinformatics and computational chemistry, the field of Nanoinformatics is relatively new\cite{maojoNanoinformaticsDevelopingNew2012}. It was only more recently that the capacity of computers reached a critical level to be able to process the vast amount of data that research in nanotechnology generates, without the necessity of resorting to high performance computing and its associated costs and limitations. The lowering of the performance barrier, the preexistence of a large knowledge base (coming from fields such as bioinformatics and cheminfornatics) and the application of computational techniques (both established and novel) have provided nanoinformatics with substantial impetus\cite{barnardNanoinformaticsBigChallenges2019}. It has enabled the adoption of simulation and modeling techniques that help fill the gap where traditional experimental techniques (such as those mentioned above) could not properly cover. Taking from advanced areas of Physics, these techniques can describe nanometric systems down to the atomic scale, providing insight into its physicochemical properties, responding for up to 30\% of the spending in pharmaceutical research spending\cite{gonzalez-niloNanoinformaticsEmergingArea2011}.

\subsection{Objective}

The objective of this work is to summarize the recent work in the field of nanoinformatics, as it applies to nanobiotechnology (NBT), in terms of the application of a set of informatics techniques classified as "nanoinformatics", as published in indexed journals, using the term as a keyword for searches.

\section{Method}

To collect the works to be reviewed, a search for the term “nanoinformatics” in the topic field was conducted in the Web of Science Portal. Since the field is relatively new and not well-defined, there is no MeSH keyword associated with it. 

The resulting 131 hits were then filtered using the criteria presented in the table below, with the respective count of articles remaining after each filter's pass.

\begin{table}[H] 
	\caption{Number of articles selected for review after passing each of the corresponding criteria.\label{tab:numeros}}
	\newcolumntype{C}{>{\centering\arraybackslash}X}
	\begin{tabularx}{\textwidth}{CC}
	\toprule
	\textbf{Filter}	& \textbf{Number of articles}\\
	\midrule
	Open Access Articles		& 64\\
	Research papers (excluding reviews and conference reports)		& 42\\
	Published in the past 3 years		& 8\\
	\bottomrule
	\end{tabularx}
\end{table}
	
\begin{table}[H] 
	\caption{Number of articles selected for review after passing each of the corresponding criteria.\label{tab:artigos}}
		\newcolumntype{C}{>{\centering\arraybackslash}X}
		\newcolumntype{Z}{>{\centering\arraybackslash}p{5em}}
		\begin{tabularx}{\linewidth}{ZCCZC}
		\toprule
		\textbf{Reference}	& \textbf{Number of articles} & \textbf{Journal} & \textbf{Year} & \textbf{Subject}\\
		\midrule
		\cite{varsouZetaPotentialRead2020}		& Varsou, et al & Small & 2020 & Cloud-based computing for image analysis\\
		\cite{lynchCanInChINano2020}		& Lynch, et al & Nanomaterials & 2020 & Proposal of a classification method and machine-readable specification for an unique identifier for nanomaterials\\
		\cite{mirzaeiMachineLearningTool2021}		& Mirzaei, et al & Nanomaterials & 2021 & A Machine Learning method for prediction of the properties of nanoparticles\\
		\cite{nawazNanoinformaticsApproachEvaluate2022}		& Nawaz, et al & Combinatorial Chemistry \& High Throughput Screening & 2022 & Developing nanoparticles for specific uses ab initio instead of following a trial-and-error approach\\
		\cite{pandaMolecularNanoinformaticsApproach2022}		& Panda, et al & Green Chemistry & 2022 & Biocompatibility of nanoparticles\\
		\cite{rostamiCharacterizationFolicAcidfunctionalized2022}		& Rostami \& Davarnejad & IET Nanobiotechnology & 2022 & Simulation and behavior prediction of nanomicelles using computational methods\\
		\cite{krikasModelingClearanceRetention2022a}		& Krikas, et al & Inhalation Toxicology & 2022 & A set of two computational models used to evaluate biodistribution of nanoparticles, confronted with validated data from the literature\\
		\cite{Blekos2023}		& Blekos, et al & Journal Of Cheminformatics & 2023 & Proposal of expansion of a previously published paradigm for nanomaterials classification (see ref. \cite{lynchCanInChINano2020})\\
		\bottomrule
		\end{tabularx}
\end{table}

\section{Discussion}

The articles collected can be separated into three different categories, with respect to which step of the research process the tools of nanoinformatics are most intensely applied, as listed in table \ref{tab:categories} below.

\begin{table}[H] 
	\caption{Number of articles selected for review after passing each of the corresponding criteria.\label{tab:categories}}
	\newcolumntype{C}{>{\centering\arraybackslash}X}
	\begin{tabularx}{\textwidth}{CC}
	\toprule
	\textbf{Category}	& \textbf{References}\\
	\midrule
	Data processing and analysis	& \cite{varsouZetaPotentialRead2020,pandaMolecularNanoinformaticsApproach2022,krikasModelingClearanceRetention2022a}\\
	Design of nanomaterials	& \cite{mirzaeiMachineLearningTool2021,nawazNanoinformaticsApproachEvaluate2022,rostamiCharacterizationFolicAcidfunctionalized2022}\\
	Ontology and classification of nanomaterials	& \cite{lynchCanInChINano2020,Blekos2023}\\
	\bottomrule
	\end{tabularx}
\end{table}

\subsection{Data processing and analysis}

Probably the most common area of application of informatics to any research field, this category contains the highest number of articles in the review set.  The use of computers has been a part of scientific research, in a data processing and analysis capacity, since its first beginnings, due to its capacity to handle large amounts of data and the elevated number of mathematical operations involved. 

It should come as no surprise, then, that the frontiers of computer science are represented in nanoinformatics as well, with Machine Learning techniques playing their part, as in predictive models used to characterize and extract descriptors associated with Transmission Electron Microscopy (TEM) images. This Cloud-based tool processes the submitted images through a series of filters that calculates a series of parameters that describe the geometry of the elements present in the image.

This process can then be used by the researcher to assess other properties of nanomaterials, which can then lead to a better understanding of their behavior. One such example is the evaluation of the Zeta Potential, which is related to the surface charge of the nanoparticle. Surface charge, in turn, is directly related to the agglomeration properties of the material, a characteristic that is an important aspect in the assessment of their toxicity\cite{varsouZetaPotentialRead2020}.

Another important aspect of computer science that finds use in NBT is a technique familiar to those involved in materials research, especially in Physics, such as Density Functional Theory (DFT) and Molecular Dynamics Simulations (MDS). DFT was originally based on the description for the electronic structure provided by the Thomas-Fermi model, but was formally proposed in 1964\cite{hohenbergInhomogeneousElectronGas1964}. It describes the electron density of a material in terms of functionals instead of the individual charged particles, and found widespread usage in the field of Solid State Physics. Developments in the 1990s led to its adaptation to applications in chemistry, providing a better computational cost to alternative approaches, but to this day this method offers challenges when dealing with more complicated systems\cite{vondrasekUnexpectedlyStrongEnergy2005}, remaining as an active research topic. MDS, on the other hand, approaches the calculation of the interaction of a set of particles with a liquid such as water by approaching the behavior of individual atoms and molecules through the calculation of interactions on set time intervals. It traces its origins back to the Monte Carlo methods used in Los Alamos Laboratory and has found more extensive use in biochemistry and biophysics than DFT. These methods find use in the solution and description of the composition and structural properties of nanoparticles, as well as their behavior in solution. The results obtained from these calculations are then used in the assessment of biocompatibility of such nanoparticles and their participation in natural processes in the organism, such as steatosis and apoptosis\cite{pandaMolecularNanoinformaticsApproach2022}.

A more traditional approach can be built by the description of the behavior of nanoparticles through differential equations. The clearance, retention and translocation of particles through different regions of the body can be described by the equilibrium between these equations, that are combined into a single model, bound by certain parameters that control the loadings of (in this case) gold at each point. The optimization of these parameters can be done in a variety of ways that involve the minimization of a certain quantity that is calculated from their values and from experimental data\cite{krikasModelingClearanceRetention2022a}.

\subsection{Design of nanomaterials}

One of the ways in which nanomaterials can be designed is the direct approach - individual placement of atoms at the desired positions, through the usage of specially designed software that can then calculate the bonding energy and viability of the proposed structure, as well as obtain the equilibrium state of the desired molecule (if any are possible). These proposed structures can then be included in MDS, as a pathway to ascertain their viability and synthesis pathways, based on the preliminary results\cite{rostamiCharacterizationFolicAcidfunctionalized2022}. Carbon nanotubes can be designed using a similar approach, with applications to Alzheimer's Disease\cite{nawazNanoinformaticsApproachEvaluate2022}.

An alternative way to plan ahead before actual synthesis of nanomaterials is the application of Machine Learning. This approach takes advantage of the already existing published experimental data in the construction of a model that calculates correlations between certain properties of the nanomaterials and their characteristics. After testing and validation, this model can then be used in the prediction of the properties of new materials, through analysis of their expected parameters\cite{mirzaeiMachineLearningTool2021}.

\subsection{Ontology and classification of nanomaterials}

The vastly wide range of properties that nanomaterials can exhibit demands some kind of classification method. The number of areas where they find application, allied to the different mechanisms through which change is effected, makes it a necessity to classify these materials in order to properly compare their utility and effectiveness. Therefore, it is a sign of the rapid development of the field of nanoinformatics that attempts at defining an ontology and classification methods have already been made. Such a proposal includes a hierarchy, composed of double descriptors related to categories and nanometric properties of materials, organized into successive levels of complexity, from basic chemical components to specific surface ligands present in the material.

Such a proposal, based on the IUPAC International Chemical Identifier (InChI), seeks to consolidate the set of properties presented by nanomaterials that should be used to categorize and identify them. Not only does this have the objective of uniformizing the identification of nanomaterials, but also seeks to provide ways for the exchange of data, as well as a possible framework for regulation of the synthesis and usage of nanomaterials\cite{lynchCanInChINano2020}. An update on that proposal discusses its objectives in terms of a standardization of language that is able to accurately describe materials, being able to encode the relevant information relevant to identification and capacity to be extended and/or future-proofed\cite{Blekos2023}.

\section{Conclusions}

Nanoinformatics is a novel, still emerging area, born from the marriage of advanced theoretical models from Physics, properties arising from the Chemistry of nanoparticles, the advancement of Computer Science, and the highly prolific field of Nanobiotechnology.

Recent works in this area demonstrate the potential of the interdisciplinary efforts from scientists from diverse fields, showing promising advances with sights to revolutionize the mechanisms available for facing an ever-growing number of issues.

Although it may seem precocious to start creating classification paradigms, it should be desirable that a systematization of language to describe the properties and behavior of nanoparticles, including attempts ate future-proofing such a system, could serve to enhance the ways new nanomaterials are planned, synthesized and characterized, further advancing research in NBT.

Care must be taken, as well, when determining criteria for the validity of NBT-based approaches for diagnosis and treatment; dated paradigms such as Lipinski's rule of five16 should be avoided, with widespread adoption of criteria that are less empirical, and more objective. In a similar way, the adoption of novel computational techniques such as artificial intelligence, machine learning and deep learning must also be approached with cautious optimism, so as not to fall into misleading conclusions.

Finally, the evolution of experimental techniques must also be followed closely; the improvement of computational techniques effects change in the capabilities of those approaches as well. Techniques such as Small-Angle Scattering, for example, have great potential for providing deep insight into nanomaterials of all kinds.

\vspace{6pt} 

\section{References}
\bibliographystyle{mdpi}
\bibliography{main.bib}

\end{document}